\documentclass[aps,prb,twocolumn,groupedaddress]{revtex4}%
\usepackage{amsmath}
\usepackage{graphics}
\usepackage{graphicx}
\usepackage{epsfig}
\usepackage{subfigure}
\usepackage{amssymb}
\usepackage{amsmath}
\usepackage{amsfonts}%

\begin{document}

\title{Quantum criticality in inter-band superconductors}
\author{Aline \surname{Ramires}, Mucio A. \surname{Continentino}}
\affiliation{Centro Brasileiro de Pesquisas F\'{\i}sicas, \\
Rua Dr. Xavier Sigaud, 150 - Urca, 22290-180,  Rio de Janeiro, RJ, Brazil}

\email{mucio@cbpf.br}


\begin{abstract}

In fermionic systems with different types of quasi-particles, attractive interactions can give rise to exotic superconducting states, as pair density wave (PDW) superconductivity and breached pairing.  In the last years the search for these new types of ground states in cold atom  and in metallic systems has been intense. 
In the case of  metals the different quasi-particles may be the up and down spin bands in an external magnetic field or bands arising from distinct
atomic orbitals that coexist at a common Fermi surface. These systems present a complex phase diagram as a function of 
the difference between the Fermi wave-vectors of the different bands. This can be controlled by external means, varying the density in the two-component cold atom system or, in a metal, by applying an external magnetic field or pressure. Here we study the 
zero temperature instability
of the normal  system as the Fermi wave-vectors mismatch of the quasi-particles (bands)
is reduced and find a second order quantum phase transition to a 
PDW superconducting state. From the nature of  the quantum critical fluctuations close to the superconducting quantum critical point (SQCP), we obtain its dynamic critical exponent. It turns out to be $z=2$ and this allows to fully characterize the SQCP for dimensions $d \ge 2$.

\end{abstract}

\maketitle

In strongly correlated materials superconductivity can be suppressed in different
ways. Most commonly, this is accomplished  by an external magnetic field, applied
pressure or doping%
\cite{leticie, loh, bianchi, mgb2, chanclog, demuer, flouquet}. However, the
point in the phase diagram where the critical temperature $T_{c}$
vanishes as a function of the external parameters is not necessarily
associated with a SQCP. In the case of
superconductivity induced by antiferromagnetic fluctuations due to
the proximity of an antiferromagnetic quantum critical point (AFQCP)\cite{steglich}, as the system
moves away from the AFQCP, these fluctuations change from
attractive to repulsive and superconductivity just fades away \cite%
{muciojpsj}. 

Inter-band superconductivity is a long standing problem in condensed matter physics \cite{suhl,fflo}. It can be realized, for example, by applying an external magnetic field in a metallic system. The field splits the band and the phase diagram as a function of the mismatch of the Fermi wave-vectors of the up and down spin bands can be investigated. It also occurs for the case of superconductivity mediated by valence fluctuations, where the most relevant correlations are due to inter-band pairing\cite{miyake}. In multi-band metals, pressure can be used to change the hybridization and the mismatch of the Fermi wave-vectors of the different bands coexisting at the common Fermi surface \cite{Padilha}. More recently this problem has received much attention due to the possibility of investigating it in cold atom systems. In this case the attractive  interaction between two fermionic systems with different Fermi wave-vectors can be tuned by Feshbach resonance\cite{caldas}. The Fermi wave-vector mismatch $\delta k_F$ in this case is controlled by varying the density of the atoms and this allows the phase diagram to be fully explored. In the core of neutron stars, up an down quarks, in different numbers, with attractive interactions may give rise to color  {\it inter-band} superconductivity \cite{caldas,Padilha}. We neglect  here orbital effects. Whenever a {\it magnetic field} is  mentioned, it is to be considered as an external agent whose only effect is to produce the mismatch of the Fermi wave-vectors. While avoiding complications related to the vortices \cite{15},  the present approach is relevant for many interesting problems  \cite{caldas}  as pointed out above.

The zero temperature phase diagram for inter-band superconductors with s-wave pairing has been established  using mean-field calculations\cite{fflo}.  As the Fermi wave-vector mismatch increases, there is a first order phase transition from a BCS-like state to an FFLO or PDW state. As the mismatch further increases there is a continuous transition  to a normal metallic phase\cite{fflo}. As new systems are discovered and the possibility of finding experimentally new exotic PDW phases increases, it is important to fully understand the nature of the normal to PDW transition beyond the mean-field approximation. Here we investigate this transition using a new approach. It is based on a
perturbation theory for retarded and advanced Green's functions\cite%
{tyablikov}.  We relate
the appearance of superconductivity in the multi-band system to the
divergence of a generalized susceptibility\cite{Thouless} $\chi(q, \omega)$,  like in  the Stoner
criterion for ferromagnetism.

We start in the normal phase and calculate the response of the system to a frequency and wave-vector
dependent {\it fictitious} external field  which couples to the superconducting
order parameter \cite{cote}. For simplicity we consider here the case of s-wave
superconductivity.  

In our approach, at the level of the RPA approximation presented here, we have to calculate
only single-particle Green's functions. This is different and simpler than
the usual linear response theory, which relates the response of the system to
two-particles Green's functions \cite{note}. 

At zero temperature, as the Fermi wave-vectors mismatch is reduced, we find a divergence of the static  part of
the susceptibility at a finite wave-vector $q=q_0$. This occurs at  the critical mean-field value of
the mismatch which destroys pair density wave superconductivity. This divergence implies
that even in zero \textit{field} the system can have a finite inhomogeneous
superconducting order parameter. The condition for the superconducting
instability can be expressed in the form of a Stoner-like criterion, $1-U
\chi_0(q=q_0, \delta k_{F})=0$. This determines either a
critical value of the attractive interaction $U_c$ above which the
system is superconductor or a critical mismatch below which
superconductivity sets in. The coupling  of the order parameter to the electronic degrees of freedom determines the frequency dependence of the generalized susceptibility and the nature of the quantum fluctuations, i.e., the dynamic critical
exponent $z$ \cite{hertz}, at the PDW SQCP. The theory is equivalent to a quantum Gaussian
approach and since the dynamic exponent turns out to be $z=2$, it yields the
correct description of the SQCP for dimensions $d \ge 2$.

We start with the following Hamiltonian describing a two-band system
with inter-band attractive interactions,
\begin{equation}  \label{HamiltonianoInter}
\mathcal{H}_0=\sum_{i,j}t^a_{ij}a_{i}^{\dagger}a_{j}
+\sum_{i,j}t^b_{ij}b_{i}^{\dagger}b_{j}-U \sum_{i}n^a_{i}n^b_{i}.
\end{equation}

For simplicity, we omitted spin indexes
\cite{caldas}. The bands $a$ and $b$ can be the up and down spin bands of a metal polarized by an external
magnetic field $h$\cite{fflo}, different types of atoms or the hybridized bands of a multi-band metal\cite{Padilha}. The
interaction $U$ is an attractive interaction between the fermions in
different bands. We calculate the response of the normal two-band system
to a wave-vector and frequency dependent
fictitious field that couples to the superconducting order parameter
of interest. in this case,
\begin{equation}
\mathcal{H}_{1}=-g\sum_{i}e^{i\mathbf{q}\cdot\mathbf{r}_{\mathbf{i}%
}}e^{i\omega _{0}t}(a_{i}^{\dagger }b_{i}^{\dagger }+a_{i}b_{i}),
\label{Hamiltoniano2}
\end{equation}
where the frequency $\omega_{0}$ has a small positive imaginary part
to guarantee the adiabatic switching on of the field. The response
of the system to the fictitious field, will be obtained using
perturbation theory for the retarded and advanced Green's
functions\cite{tyablikov}. We start in the normal phase where the
superconducting order parameter is zero in the absence of the
\textit{external field} $g$. We split the Green's functions, normal
and anomalous, in two contributions. The first of order zero and the
second of first order in the field $g$. For the anomalous Green's functions $\ll a_{i\sigma }^{\dagger }|b_{j\sigma ^{\prime }}^{\dagger }\gg$, for example, we write,
\[
{\ll a_{i\sigma }^{\dagger }|b_{j\sigma ^{\prime }}^{\dagger }\gg }%
\rightarrow {\ll a_{i\sigma }^{\dagger }|b_{j\sigma ^{\prime }}^{\dagger
}\gg }^{(0)}+{\ll a_{i\sigma }^{\dagger }|b_{j\sigma ^{\prime }}^{\dagger
}\gg }^{(1)}.
\]

In the normal phase and in the absence of the fictitious field, the relevant
zero order Green's functions can be easily calculated,
\[
G_k^{aa} (\omega)={\ll a_{k}|a_{k^{\prime }}^{\dagger}\gg }^{(0)}_{\omega}= \frac{\delta_{k,k^{\prime}}}{2\pi(\omega-\epsilon_k^a)},
\]%
\[
G_k^{bb}(\omega)={\ll b_{k}|b_{k^{\prime }}^{\dagger }\gg }%
^{(0)}_{\omega}=\frac{\delta_{k,k^{\prime}}}{2\pi(\omega-\epsilon_k^b)},
\]%
\[
{\ll a_{k}|b_{k^{\prime }}^{\dagger }\gg }%
^{(0)}_{\omega}=0,
\]%
\[
{\ll a_{k}^{\dagger }|a_{k^{\prime }}^{\dagger }\gg }%
^{(0)}_{\omega}=0,
\]%
\[
{\ll b_{k}^{\dagger }|b_{k^{\prime }}^{\dagger }\gg }%
^{(0)}_{\omega}=0,
\]
and
\[
{\ll a_{k}^{\dagger }|b_{k^{\prime }}^{\dagger }\gg }%
^{(0)}_{\omega}={\ll b_{k}^{\dagger }|a_{k^{\prime }}^{\dagger }\gg }%
^{(0)}_{\omega}=0,
\]
where $\epsilon_k^{\alpha} =\sum_j t_{ij}^{\alpha} e^{i\mathbf{k}.(\mathbf{r}_j- \mathbf{r}_i)}$, ($\alpha=a,b$). 
For the first order Green's functions we get,
\begin{eqnarray}
(\omega\!-\!\varepsilon_k^a){\ll \!\!a_{k
}|a^{\dagger}_{k^{\prime}}\!\!\gg}\!
_{\omega}^{(1)} \!=\! (U\delta \Delta^{ab}_{k}\! + \!g) {\ll \!\!b^{\dagger}_{k\! + \!q}|
a_{k^{\prime}}^{\dagger}\!\!\gg}^{(0)}_{\!\!\omega\!+ \! \omega_0},\nonumber \\
(\omega\!  - \!\varepsilon_k^b){\ll \! \! b_{k
}|b^{\dagger}_{k^{\prime}}\!\!\gg}%
_{\omega}^{(1)} \!  = \! (U\delta \Delta^{ab}_{k}\!  - \!g) {\ll \! \! a^{\dagger}_{k\! + \!q}|
b_{k^{\prime}}^{\dagger}\!\!\gg}^{(0)}_{\omega\! + \!\omega_0},\nonumber \\
(\omega\!  - \!\varepsilon_k^a){\ll \! \! a_{k
}|b^{\dagger}_{k^{\prime}}\!\!\gg}%
_{\omega}^{(1)} \!  = \! (U\delta \Delta^{ab}_{k}\! + \!g) {\ll \! \! b^{\dagger}_{k\! + \!q}|
b_{k^{\prime}}^{\dagger}\!\!\gg}^{(0)}_{\omega\! + \!\omega_0},\nonumber \\
(\omega\! + \!\varepsilon_k^a){\ll \! \! a^{\dagger}_{k
}|a^{\dagger}_{k^{\prime}}\!\!\gg}%
_{\omega}^{(1)} \! \! = \! \!  - (U\delta \Delta^{ab}_{k}\! + \!g) {\ll \! \! b_{k\! + \!q}|
a_{k^{\prime}}^{\dagger}\!\!\gg}^{(0)}_{\omega\! + \!\omega_0},\nonumber \\
(\omega\! + \!\varepsilon_k^b){\ll \! \! b^{\dagger}_{k
}|b^{\dagger}_{k^{\prime}}\!\!\gg}%
_{\omega}^{(1)} \!  \!= \!  - (U\delta \Delta^{ab}_{k}\!  - \!g) {\ll \! \! a_{k\! + \!q}|
b_{k^{\prime}}^{\dagger}\!\!\gg}^{(0)}_{\omega\! + \!\omega_0},\nonumber \\
(\omega\! + \!\varepsilon_k^a){\ll \! \! a^{\dagger}_{k
}|b^{\dagger}_{k^{\prime}}\!\!\gg}%
_{\omega}^{(1)} \! \! = \! \!  - (U\delta \Delta^{ab}_{k}\! + \!g) {\ll \! \! b_{k\! + \!q}|
b_{k^{\prime}}^{\dagger}\!\!\gg}^{(0)}_{\omega\! + \!\omega_0}.
\end{eqnarray}

The first three first order propagators are written in terms
of zero order  propagators associated with superconductivity. Since we are analyzing the system in the normal
state, these anomalous zero order propagators vanish and consequently the first
order ones also vanish. The three last first order propagators are written
in terms of conventional zero order propagators, but just the last one
is different from zero. It is given by,
\begin{equation}
{\ll a_{k}^{\dagger }|b^{\dagger}_{k^{\prime}}\gg }%
_{\omega }^{(1)}\!=\!-\delta_{k+q,k^{\prime}} \frac{\left( U\delta \Delta^{ab}_{k}+g\right)}{2\pi(\omega +\epsilon
_{k}^{a})(\omega+\omega_0-\epsilon_{k+q}^b)}.
\end{equation}

We use the fluctuation-dissipation theorem%
\cite{tyablikov} to obtain the correlation function $\delta \Delta ^{aa}$
from the above anomalous Green's functions, we get,
\[
\delta \Delta ^{ab}=\sum_{k}F_{\omega }\left\{ {\ll a_{k}^{\dagger
}|b_{k^{\prime }}^{\dagger }\gg }_{\omega }^{(1)}\right\},
\]%
where  $F_{\omega }\{G(\omega )\}=-\int d\omega f(\omega )[G(\omega
+i\epsilon )-G(\omega -i\epsilon )]$ is the statistical average of the
discontinuity of the Green's functions $G(\omega )$ on the real axis\cite%
{tyablikov}. The function $f(\omega )$ is the Fermi-Dirac distribution.

We define the generalized inter-band susceptibility by, $\chi_0^{ab}(q, \omega) =2 \pi \sum_k F_{\omega} \left\{G_k^{aa}(-\omega) G_{k+q}^{bb}(\omega + \omega_0)\right\}$ or,
\begin{equation} \label{chi0}
\chi _{0}^{ab}(q,\omega )=\frac{-1}{2\pi }\sum_{k}F_{\omega }\left[ \frac{1}{%
(\omega +\varepsilon _{k}^{a})}\frac{1}{(\omega +\omega _{0}-\varepsilon
_{k-q}^{b})}\right].
\end{equation}
The superconducting response to the external field
is obtained in the form,
\begin{equation}\label{gsinter}
\delta \Delta ^{ab}=\frac{\chi _{0}^{ab}(q,\omega )}{1-U\chi
_{0}^{ab}(q,\omega )}g.
\end{equation}%
Finally, the susceptibility $\chi _{0}^{ab}(q,\omega )$, (Eq. \ref{chi0}),
is given by,
\begin{equation} \label{chi}
\chi _{0}^{ab}(2q,\omega )=\frac{1}{2\pi }\sum_{k}\frac{1-f(\epsilon
_{k-q}^{a})-f(\epsilon _{k+q}^{b})}{\epsilon _{k-q}^{a}+\epsilon
_{k+q}^{b}-\omega _{0}}.
\end{equation}

Let us consider the case of a metal polarized by an external magnetic field $h$. In this case the $a$ and $b$-bands are given by the  dispersion relation of the up and down spin bands, respectively,
\[
\epsilon _{k}^{a,b}=\frac{\hbar ^{2}(k^{2}-k_{F}^{2})}{2m} \mp h,
\]%
where $k_F$ is the original Fermi wave-vector of the unpolarized band ($h=0$). The mismatch of the new Fermi wave-vectors is given by, $\delta
k_{F}=k_{F}^{a }-k_{F}^{b }=h/v_{F}$, where $v_{F}$
is the Fermi velocity. At zero temperature, the generalized
susceptibility
$\chi_{0}^{ab}(2q,\omega _{0}=0)$ can be calculated and  we get,
\begin{eqnarray} \label{chi2}
&\chi _{0}^{ab}(2q,0) =  \rho \left\{ 1\!-\! \ln \left(\!
\frac{h}{v_{F} k_{C}}\right) \!-\!\frac{1}{2}\left[ \frac{1}{
\overline{q}}\ln \left( \frac{\overline{q}+1}{\overline{q}-1}\right) + \right. \right. \nonumber   \\
&\left. \left. \ln (\overline{q}^{2}\!-\!1)\right] \right\} ,
\end{eqnarray}
where $\overline{q}=v_{F}q/h=q/(k_{F}^{a}-k_{F}^{b})$ and $\rho=3/(8 \pi^3 E_F)$ is the density of states of the
unpolarized system. The quantum critical point separating the normal
metal from the superconducting PDW phase is obtained from the condition, $%
1-U\chi _{0}^{ab}(2q,0)=0$, which can be written as,
\begin{equation}\label{stoner}
U \rho\left\{\! 1\!-\!\ln \left(
\frac{h}{\Delta_{0}}\right)\! -\! \frac{1}{2}\left[ \frac{1}{%
\overline{q}}\ln \left( \frac{\overline{q}+1}{\overline{q}-1}\right) \!+\!\ln (%
\overline{q}^{2}-1)\right] \right\} \!=\!0,
\end{equation}
where $\Delta _{0}=v_{F}k_{C}\exp (-1/\rho U)$. Notice that a PDW state is only possible for $\overline{q}>1$. The equation above determines the critical field
$h_{N}$ (or mismatch) for a fixed value of \ $\overline{q}$. This is
given by,
\begin{equation}\label{hn}
\frac{h_{N}}{\Delta _{0}}=\frac{e}{(1+\overline{q})}\left(
\frac{\overline{q}+1}{\overline{q}-1}\right)
^{\frac{\overline{q}-1}{2\overline{q}}}.
\end{equation}
\begin{figure}[tbp]
\begin{center}
\includegraphics[width=0.65\linewidth,
  keepaspectratio]{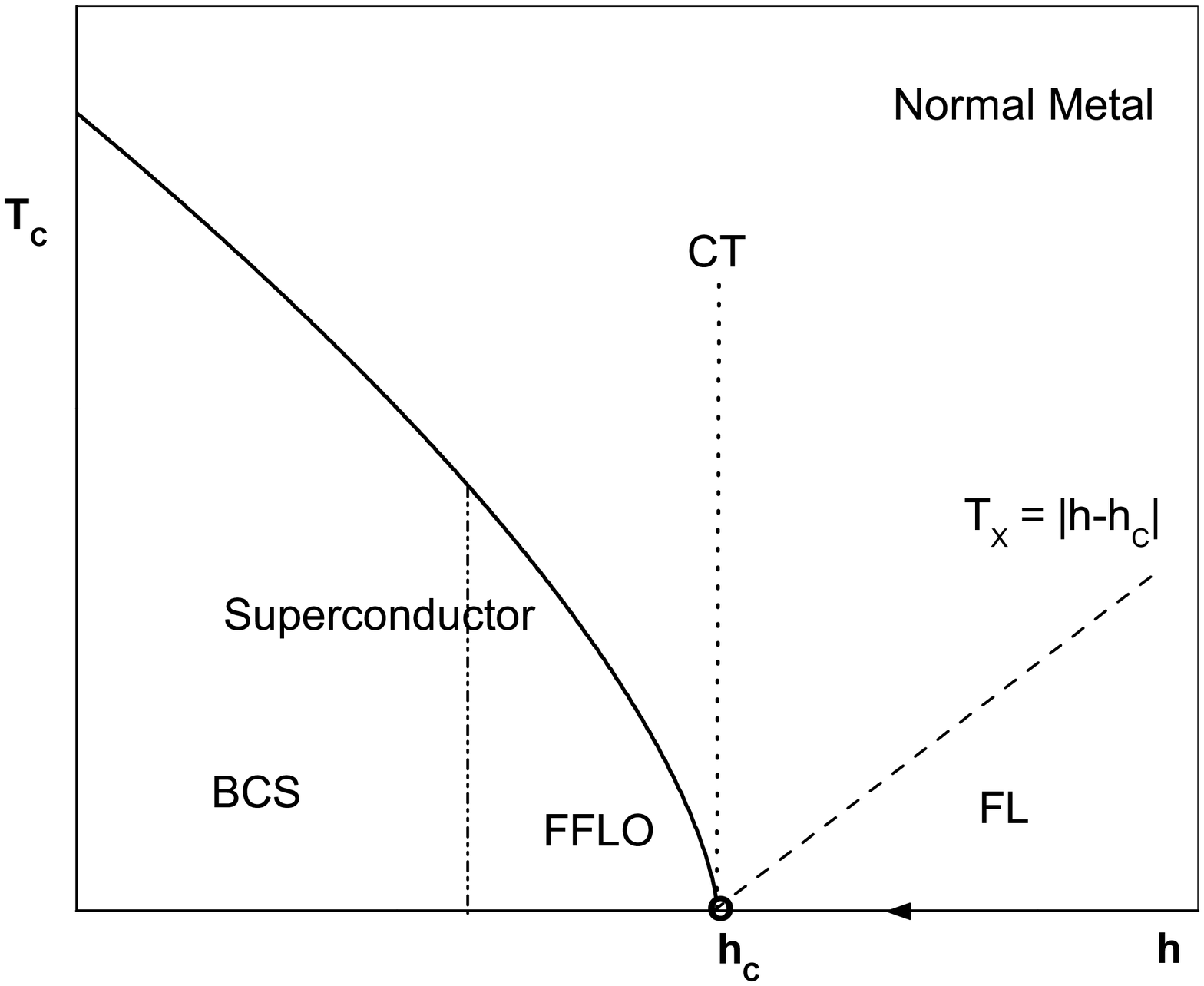}
\end{center}
\caption{Schematic $T_{c} \times h$ phase diagram for the inter-band superconductor. At zero temperature, as the external magnetic
field or Fermi wave-vectors mismatch is reduced (indicated by the arrow), the normal metal becomes unstable to an inhomogeneous
superconducting phase. This is signaled by the divergence of the
$q$-dependent generalized susceptibility. The FFLO superconductor quantum critical point at $h_{C}$ has a dynamic exponent $z=2$ and this allows to determine the behavior of the thermodynamic quantities along the critical trajectory (CT) as given in the text. As the external field is further reduced there is a first order transition from the FFLO state to a BCS superconductor\cite{Padilha} which is not studied here. The crossover line $T_{\times}$ marks the onset of Fermi liquid behavior with decreasing temperature.} \label{fig1}
\end{figure}

The maximum value of $h_{N}$ for which the instability occurs is denoted by $h_{C}$. It is easily obtained by
differentiating $h_{N}(\overline{q})$ with respect to $\overline{q}$. From
the equation ($\partial \ln h_{N}/\partial \overline{q})_{\overline{q}=%
\overline{q}_{C}}=0$, \ we find that:
\[
\overline{q}_{C}=\frac{1}{2}\ln \left( \frac{\overline{q}_{C}+1}{\overline{q}_{C}-1}%
\right) .
\]
This gives $\overline{q}_{C}\cong 1.2$, that substituted in Eq.
\ref{hn} for $h_{N}$ yields $h_{C}=h_{N}(\overline{q}_{C})\cong
1.5\Delta _{0}$. This agrees with the result of Takada and Izuyama\cite{takada}  for the vanishing of the FFLO phase starting from this phase.

As discussed in Ref.\cite{Padilha} an FFLO phase can also be induced by  hybridization ($V$).
The results above can be easily extended for the case of hybrid bands\cite{Padilha}. The Fermi wave-vectors mismatch is  given by,  $k_{F}^{a} - k_{F}^{b}=4V/v_{F}(1+\alpha)$, where $\alpha <1$ is the ratio of the effective masses of the quasi-particles. The mismatch is proportional to the hybridization  $V$ (or pressure), which now plays the role of the magnetic field.

In order to study the fluctuations close to the PDW SQCP we have to expand
Eq. \ref{chi} for small frequencies $\omega_{0}$, $h\cong h_{C}$ and $q\cong q_{C}$.
At the same level of approximation for $q$ of Eq. \ref{chi2}, we get for the denominator of the generalized susceptibility in Eq. \ref{gsinter},
\begin{equation}
1-U\chi^{ab}_{0} (2q, \omega_{0})\!=\!U \rho \left[ \frac{h\!-\!h_{C}}{h_{C}} \!+i \frac{\omega_{0}}{v_{F}q_{C}}\! +\!\frac{1}{q_{C}^2\!-\!1} (q\!-\!q_{C})^2\! \right].
\end{equation}
The coupling of the superconductor order parameter to the electronic degrees of freedom gives rise to Landau damping of the superconductor quantum fluctuations. These modes are purely evanescent with an imaginary dispersion relation.

As in Hertz approach\cite{hertz} to quantum phase transitions, we can construct from the dynamical susceptibility a quantum Gaussian action for this problem\cite{mac}, which can be written as, 
\begin{equation}\label{gaussian}
\mathcal{S}=\int d^dq \int d\omega \left[g + \omega + q^2\right] |\psi(q, \omega)|^2,
\end{equation}
where $g=h-h_C$ measures the distance to the PDW SQCP,  $\psi(q, \omega)$ is the PDW superconductor order parameter and $q-q_C \rightarrow q$.  The scaling properties of the free energy associated with this action allows to identify the dynamic exponent of the SQCP at which the FFLO instability occurs. This is easily found as $z=2$. 
The knowledge of the dynamic exponent and the quantum hyperscaling relation\cite{mac,pre}, $2-\alpha=\nu(d+z)$, allows to fully characterize the universality class of the FFLO quantum phase transition. Since $d_{eff}=d+z$, for $d=2$ and $d=3$, the critical exponents are Gaussian with possible logarithmic corrections for $d=2$. The reason is that for $d\ge 2$, interaction between the fluctuations are irrelevant in the renormalization group sense and the Gaussian action, Eq. \ref{gaussian}, gives the correct description of the quantum phase transition\cite{mac}.

Next step in the calculations is to extend the results to finite temperatures and particularly to obtain the correction to the mean-field value of the shift exponent  $\psi$ of the critical line at very low temperatures close to the PDW SQCP. This can be carried out in different ways\cite{psi}, such that, $T_C(h) \propto |h-h_C|^{1/\psi}$. It turns out that \cite{psi}  $\psi=z/(d+z-2)$ for $d_{eff} >4$ and can be determined from our knowledge of the dynamic exponent, which yields $\psi=2/d$. In the same way, scaling\cite{mac} gives the contribution of quantum critical fluctuations to the specific heat, $C/T \propto T^{(d-2)/2}$, along the critical trajectory (CT) shown in Fig. \ref{fig1}. Also shown in this figure is the crossover line\cite{mac}, $T_{\times} \propto |h-h_{C}|$, below which the system enters a Fermi liquid regime. We have used that $\nu z=1$ for $d \ge 2$.

We have studied the zero temperature instability of a normal two-band system with inter-band attractive
interactions as the Fermi wave-vectors mismatch is reduced. The model describes a superconductor in
an external field in the absence of orbital effects, a cold atom system with two atomic species  or a two-band metal with mixing tuned by pressure. We find that  as the mismatch of the Fermi
wave-vectors is reduced, the
system becomes unstable to an inhomogeneous, FFLO-like,
superconducting phase characterized by a wave-vector
$\overline{q}=\overline{q}_C$.  
We have introduced a new method to deal with systems coupled
to a space and time dependent \textit{external field} and respond
adiabatically to this perturbation. The appearance of superconductivity
is related to the divergence of a static generalized susceptibility for $%
\overline{q}=\overline{q}_C$ at a SQCP. 
The results obtained
are valid to first order in the perturbation and coincide with those
of linear response. However, in our approach the basic
elements to be calculated are single particle Green's functions and not the usual two
particle propagators of linear response theory. 
The present theory extends the mean-field  treatment to include fluctuations close to this SQCP.  We have obtained the dynamical critical exponent at the PDW SQCP and from that we have identified the
universality class of this quantum critical point for $d \ge 2$. This allows several predictions to
be made for the thermodynamic and transport properties  close to the
quantum phase transition and for the shape of the critical line for dimensions $d\ge 2$. 

We would like to thank Igor Padilha for useful discussions and the Brazilian agencies, CNPq and FAPERJ for partial financial support.


\begin{thebibliography}{99}
\bibitem{note} In practice these are decoupled as products of single particles Greens functions \cite{cote}.
\end{thebibliography}
\end{document}